\documentclass[a4paper]{jpconf}
\usepackage{graphicx}
\begin{document}
\title{Non-singular Universes \`{a} la Palatini}

\author{Gonzalo J. Olmo\footnote{Published under license in Journal of Physics: Conference Series by IOP Publishing Ltd.}}

\address{Departamento de F\'{i}sica Te\'{o}rica and IFIC, Centro Mixto Universidad de Valencia - CSIC.
  Facultad de F\'{i}sica, Universidad de Valencia, Burjassot-46100, Valencia, Spain  and \\
  Instituto de Estructura de la Materia, CSIC, Serrano 121, 28006 Madrid, Spain}

\ead{gonzalo.olmo@uv.es}

\begin{abstract}
It has recently been shown that f(R) theories formulated 
in the  Palatini variational formalism 
are able to avoid the big bang singularity yielding 
instead a bouncing solution. The mechanism responsible 
for this behavior is similar to that observed in the 
effective dynamics of loop quantum cosmology and an f(R) 
theory exactly reproducing that dynamics has been found. 
I will show here that considering more general actions, with 
quadratic contributions of the Ricci tensor, results in 
a much richer phenomenology that
yields bouncing solutions even in anisotropic (Bianchi 
I) scenarios. Some implications of these results are discussed. 
\end{abstract}

\section{Introduction.}
 
 In the last years it has been shown that though many $f(R)$ models have the ability to produce late-time cosmic acceleration and fit well the background expansion history, they are not in quantitative agreement with the structure and evolution of cosmic inhomogeneities (see \cite{f(R)} and references therein). Additionally, the fact that matter is concentrated in discrete structures like atoms causes the modified dynamics to manifest also in laboratory experiments\cite{Olmo2007b,Olmo-2008a,LMS}, which confirms earlier suspicions on the viability of such models according to their corresponding Newtonian and post-Newtonian properties \cite{Olmo2005}. 
This is a very disturbing aspect of the models with infrared corrections, which demands the consideration of a microscopic description of the sources and prevents the use of macroscopic, averaged representations of the matter. A careful analysis of this point put forward the existence of non-perturbative effects induced by the Palatini dynamics in a number of contexts \cite{Flanagan-2004a,Olmo2007b,Sot08a,Olmo-2008a}. Though this certainly is an undesired property of infrared-corrected models, it could become a very useful tool for models with corrections at high curvatures. Can we construct singularity-free cosmological models that recover GR at low curvatures using the non-perturbative properties of Palatini theories? As we report here, ultraviolet-corrected Palatini models turn out to be very efficient at removing the big bang cosmic singularity in various situations of interest.

\section{Non-singular $f(R)$ cosmologies.}

The recent interest in the dynamics of the early-universe in Palatini theories arose, in part, from the observation that the effective equations of loop quantum cosmology \cite{LQC} (LQC) could be exactly reproduced by a Palatini $f(R)$ Lagrangian \cite{OS-2009}. In LQC, non-perturbative quantum gravity effects lead to the resolution of the big bang singularity by a quantum bounce without introducing any new degrees of freedom. Though fundamentally discrete, the theory admits a continuum description in terms of an effective Hamiltonian that in the case of a homogeneous and isotropic universe filled with a massless scalar field leads to the following Friedmann equation
\begin{equation}\label{eq:LQC}
3H^2={8\pi G}\rho\left(1-\frac{\rho}{\rho_{crit}}\right) \ ,
\end{equation}
where $\rho_{crit}\approx 0.41\rho_{Planck}$. At low densities, $\rho/\rho_{crit}\ll 1$, the background dynamics is the same as in GR, whereas at densities of order $\rho_{crit}$ the non-linear new matter contribution forces the vanishing of $H^2$ and hence a cosmic bounce. This singularity avoidance seems to be a generic feature of loop-quantized universes \cite{Param09}. \\
Palatini $f(R)$ theories share with LQC the fact that the modified dynamics that they produce is not due to the existence of new dynamical degrees of freedom but rather to non-linear effects produced by the matter sources, which contrasts with other approaches to quantum gravity and to modified gravity. This allows to relate Eq.(\ref{eq:LQC}) with the corresponding $f(R)$ equation
\begin{equation}\label{eq:H2}
3H^2=\frac{f_R\left(\kappa^2\rho+({R}f_R-f)/2\right)}{\left(f_R-\frac{12\kappa^2\rho f_{RR}}{2 ({R}f_{RR}-f_R)}\right)^2} \ .
\end{equation}
Taking into account the trace of the field equations (see \cite{f(R)} for details), which for a massless scalar becomes $R f_R -2f=2\kappa^2\rho$ and implies that $\rho=\rho({R})$, one finds that a Palatini $f(R)$ theory able to reproduce the LQC dynamics (\ref{eq:LQC}) must satisfy the differential equation
\begin{equation}
f_{RR}=-f_R\left(\frac{A f_R -B}{2({R}f_R-3f)A+{R}B}\right)
\end{equation}
where $A=\sqrt{2({R}f_R-2f)(2{R}_c-[Rf_R-2f])}$, $B=2\sqrt{{R}_cf_R(2{R} f_R-3f)}$, and ${R}_c\equiv \kappa^2\rho_c$. Imposing suitable boundary conditions, one can show \cite{OS-2009} that the function 
\begin{equation}\label{eq:f-guess}
\frac{df}{dR}=- \tanh \left(\frac{5}{103}\ln\left[\left(\frac{R}{12\mathcal{R}_c}\right)^2\right]\right)
\end{equation} 
faithfully captures the LQC dynamics from low curvatures up to the non-perturbative bouncing region.
Though the function (\ref{eq:f-guess}) implies that the LQC Lagrangian is an infinite series, which is a manifestation of the non-local properties of the quantum geometry, one can find non-singular cosmologies with a finite number of terms. In fact, a simple quadratic Lagrangian of the form $f(R)=R+R^2/R_P$ does exhibit non-singular solutions for certain equations of state \cite{BOSA-2009} depending on the sign of $R_P$. To be precise, if $R_P > 0$ the bounce occurs for sources with $w=P/\rho> 1/3$. If $R_P < 0$, then the bouncing condition is satisfied by $w < 1/3$.\\ 
 
 Besides avoiding the development of curvature singularities, bouncing cosmologies can solve the horizon problem \cite{Novello-2008}, which makes them interesting as a substitute for inflation. However, it has been found \cite{Koi-2010} that $f(R)$ models that develop a bounce when the condition $f_R=0$ is met turn out to exhibit singular behavior of inhomogeneous perturbations in a flat, dust-filled universe. Further insight on the robustness of the bounce under perturbations  was obtained in \cite{BO-2010} studying the properties of $f(R)$ theories in anisotropic spacetimes of Bianchi-I type, for which $ds^2=-dt^2+\sum_{i=1}^3 a_i^2(t)(dx^i)^2$. In space-times of this type with a perfect fluid, one can derive a number of useful analytical expressions. In particular, one finds that the expansion $\theta=\sum_i H_i$ and the shear $\sigma^2=\sum_i\left(H_i-\frac{\theta}{3}\right)^2$ are given by 
\begin{equation}\label{eq:Hubble-f(R)}
\sigma^2=\frac{\rho^{\frac{2}{1+w}}}{f_R^2}\frac{(C_{12}^2+C_{23}^2+C_{31}^2)}{3}  \ , \ 
\frac{\theta^2}{3}\left(1+\frac{3}{2}\Delta_1\right)^2=\frac{f+\kappa^2(\rho+3P)}{2f_R}+\frac{\sigma^2}{2}
\end{equation}
where the constants $C_{ij}=-C_{ji}$ set the amount and distribution of anisotropy and satisfy the constraint $C_{12}+C_{23}+C_{31}=0$. In the isotropic case, $C_{ij}=0$, one has $\sigma^2=0$ and $\theta^2=9H^2$, with $H^2$ given by Eq.(\ref{eq:H2}). Now, 
since homogeneous and isotropic bouncing universes always require the condition $f_R=0$ at the bounce \cite{BO-2010}, a glance at (\ref{eq:Hubble-f(R)})  indicates that the shear diverges as $\sim 1/f_R^2$. This shows that any isotropic $f(R)$ bouncing model will develop divergences when anisotropies are present. Moreover,  
one can check by direct calculation that the Kretschman scalar $R_{\mu\nu\sigma\rho}R^{\mu\nu\sigma\rho}=4(\sum_i(\dot H_i+H_i^2)^2+H_1^2H_2^2+H_1^2H_3^2+H_2^2H_3^2)$ diverges at least as $\sim 1/f_R^4$, which is a clear geometrical pathology and signals the presence of a physical singularity. 

\section{Nonsingular cosmologies beyond $f(R)$.}

The previous section provides reasons to believe that Palatini $f(R)$ models are not able to produce a fully satisfactory and singularity-free alternative to GR. Though the homogeneous and isotropic case greatly improves the situation with respect to GR, the existence of divergences when anisotropies and inhomogeneities are present spoils the hopes deposited on this kind of Lagrangians. To the light of these results, we explored new Palatini theories \cite{BO-2010} to determine if the introduction of new geometrical invariants in the gravitational action could avoid the problems that appear in the $f(R)$ models. We considered Lagrangians of the form $f(R,Q)$, with $Q=R_{\mu\nu}R^{\mu\nu}$. Using the particular Lagrangian 
\begin{equation}\label{eq:f(R,Q)}
f(R,R_{\mu\nu}R^{\mu\nu})=R+a\frac{R^2}{R_P}+\frac{R_{\mu\nu}R^{\mu\nu}}{R_P} \ ,
\end{equation}
it was found that completely regular bouncing solutions exist for both isotropic and anisotropic homogeneous cosmologies filled with a perfect fluid. In particular, we found that for $a<0$ the interval $0\leq w\leq 1/3$ is always included in the family of bouncing solutions, which contains the dust and radiation cases. For $a\geq 0$, the fluids yielding a non-singular evolution are restricted to $w>\frac{a}{2+3a}$, which implies that the radiation case $w=1/3$ is always nonsingular. For a detailed discussion and classification of the non-singular solutions depending on the value of the parameter $a$ and the equation of state $w$ see \cite{BO-2010}. \\
The field equations that follow from the Lagrangian (\ref{eq:f(R,Q)}) when $R_{\mu\nu}$ is assumed symmetric were derived in \cite{OSAT} and take the form 
\begin{equation}\label{eq:con-var}
f_R R_{\mu\nu}-\frac{f}{2}g_{\mu\nu}+2f_QR_{\mu\alpha}{R^\alpha}_\nu = \kappa^2 T_{\mu\nu} \ , \ \nabla_{\beta}\left[\sqrt{-g}\left(f_R g^{\mu\nu}+2f_Q R^{\mu\nu}\right)\right]=0 
\end{equation}
where $f_R\equiv \partial_R f$ and $f_Q\equiv \partial_Q f$. The connection equation in (\ref{eq:con-var}) can be solved introducing an auxiliary metric $h_{\alpha\beta}$ such that it takes the form $\nabla_{\beta}\left[\sqrt{-h} h^{\mu\nu}\right]=0$, which implies that $\Gamma^{\rho}_{\mu\lambda}$ can be written as the Levi-Civita connection of $h_{\mu\nu}$. When the matter sources are represented by a perfect fluid, $T_{\mu\nu}=(\rho+P)u_\mu u_\nu+P g_{\mu\nu} $, one can show that $h_{\mu\nu}$ and $h^{\mu\nu}$ are given by \cite{OSAT}
\begin{equation}
h_{\mu\nu}=\Omega\left( g_{\mu\nu}-\frac{\Lambda_2}{\Lambda_1-\Lambda_2} u_\mu u_\nu \right)\ , \ 
h^{\mu\nu}=\frac{1}{\Omega}\left( g^{\mu\nu}+\frac{\Lambda_2}{\Lambda_1} u^\mu u^\nu \right)
\end{equation}
where  $\lambda=\sqrt{\kappa^2 P+\frac{f}{2}+\frac{f_R^2}{8f_Q}}$, $\Lambda_1= \sqrt{2f_Q}\lambda+\frac{f_R}{2}$, $\Lambda_2= \sqrt{2f_Q}\left[\lambda\pm\sqrt{\lambda^2-\kappa^2(\rho+P)}\right]$, and $\Omega=\left[\Lambda_1(\Lambda_1-\Lambda_2)\right]^{1/2}$.
The metric field equation in (\ref{eq:con-var}) takes the following form 
\begin{equation}\label{eq:Rmn-h}
R_{\mu\nu}(h)=\frac{1}{\Lambda_1}\left[\frac{\left(f+2\kappa^2P\right)}{2\Omega}h_{\mu\nu}+\frac{\Lambda_1\kappa^2(\rho+P)}{\Lambda_1-\Lambda_2}u_{\mu}u_{\nu}\right] \ .
\end{equation}
In this expression, the functions $f, \Lambda_1$, and $\Lambda_2$ are functions of the density $\rho$ and pressure $P$. In particular, in a universe filled with radiation, for which $R=0$, the function $Q$ in our quadratic model (\ref{eq:f(R,Q)}) boils down to $Q= \frac{3R_P^2}{8}\left[1-\frac{8\kappa^2\rho}{3R_P}-\sqrt{1-\frac{16\kappa^2\rho}{3R_P}}\right] \label{eq:Q-rad}$. 
This expression recovers the GR value at low curvatures, $Q\approx 4(\kappa^2\rho)^2/3+32(\kappa^2\rho)^3/9R_P+\ldots$ but reaches a maximum $Q_{max}=3R_P^2/16$ at $\kappa^2\rho_{max}=3R_P/16$. At $\rho_{max}$ the shear also takes its maximum allowed value, namely, $\sigma^2_{max}=\sqrt{3/16}R_P^{3/2}(C_{12}^2+C_{23}^2+C_{31}^2)$, which is always finite, and the expansion vanishes producing a cosmic bounce regardless of the amount of anisotropy. \\
It should be noted that the choice of a symmetric Ricci tensor in the analysis of $f(R,Q)$ bouncing cosmologies presented above is not arbitrary. As pointed out in \cite{Vitagliano:2010pq}, the antisymmetric part of the Ricci tensor introduces new dynamical degrees of freedom in the form of a massive vector field (see also \cite{Vollick:2006uq} for a related result). If one looks for a framework suitable for the description of the effective dynamics of LQC (including anisotropies) and, more generally, of other theories of quantum geometry not involving new degrees of freedom, it seems natural to impose constraints on the spectrum of possible Lagrangians to avoid new propagating fields. In this sense, it should be noted that the $f(R,Q)$ theories discussed here are also able to reproduce \cite{Olmo-2011} other aspects of the expected phenomenology of quantum gravity at the Planck scale. In particular, without imposing any a priori phenomenological structure, the quadratic Palatini model (\ref{eq:f(R,Q)}) predicts an energy-density dependence of the metric components that closely matches the structure conjectured in models of Doubly (or Deformed) Special Relativity and Rainbow Gravity \cite{DSR-RG}. This confirms that Palatini theories represent a new and powerful framework to address different aspects of quantum gravity phenomenology.\\

Work supported by the Spanish grants FIS2008-06078-C03-02, FIS2008-06078-C03-03, and the Consolider Programme CPAN (CSD2007-00042).


\section*{References}
\begin{thebibliography}{9}

\bibitem{f(R)}
G. J. Olmo, Int.\ J.\ Mod.\ Phys.\  {\bf D}, to appear (2011), [arXiv:1101.3864 [gr-qc]];
A.~De Felice, S.~Tsujikawa, Living Rev.\ Rel.\  {\bf 13}, 3 (2010);
T.P. Sotiriou and V. Faraoni, Rev. Mod. Phys. 82, 451-497 (2010).


\bibitem{Olmo2007b}
G. J. Olmo, Phys. Rev. Lett. {\bf 98}, 061101 (2007).


\bibitem{Olmo-2008a}
G. J. Olmo, Phys. Rev. {\bf D 77}, 084021(2008).

\bibitem{LMS}
B.~Li, D.~F.~Mota and D.~J.~Shaw, Phys. Rev. {\bf D 78}, 064018 (2008);  Class.\ Quant.\ Grav.\  {\bf 26}, 055018 (2009).

\bibitem{Olmo2005}
 G.~J.~Olmo,   Phys.\ Rev.\ Lett.\  {\bf 95}, 261102 (2005);  Phys.\ Rev.\  {\bf D72}, 083505 (2005).

\bibitem{Flanagan-2004a}
E. E. Flanagan, Phys. Rev. Lett.{\bf 92}, 071101 (2004).

\bibitem{Sot08a} 
E. Barausse, T.P. Sotiriou, and J.C. Miller, {\it Class.Quant.Grav.} 25,062001(2008);
{\it Class.Quant.Grav.} 25,105008(2008). 

\bibitem{LQC}
M. Bojowald, {\it Living Rev. Rel.} {\bf 8}, 11 (2005);
A. Ashtekar, {\it Nuovo Cim.} {\bf 122 B}, 135 (2007);
A. Ashtekar, T. Pawlowski and P. Singh, {\it Phys. Rev. Lett. } {\bf 96}, 141301 (2006); {\it Phys. Rev.} {\bf D} 73,  124038 (2006);
{\it Phys. Rev. } {\bf D} 74, 084003 (2006); 
L. Szulc, W. Kaminski, J. Lewandowski, {\it Class.Quant.Grav.} {\bf 24}, 2621 (2007); 
A. Ashtekar, T. Pawlowski, P.Singh, K. Vandersloot, {\it Phys.Rev.} {\bf D} 75, 024035 (2007); 
K. Vandersloot, {\it Phys.Rev.} {\bf D} 75, 023523 (2007).

\bibitem{OS-2009}
G.J.Olmo and P.Singh, JCAP 0901, 030 (2009).

\bibitem{Param09}
  P.~Singh,
  Class.\ Quant.\ Grav.\  {\bf 26}, 125005 (2009).

\bibitem{BOSA-2009}
C.~Barragan, G.~J.~Olmo, H.~Sanchis-Alepuz,
  Phys.\ Rev.\  {\bf D80}, 024016 (2009);
C.~Barragan, G.~J.~Olmo, H.~Sanchis-Alepuz, [arXiv:1002.3919 [gr-qc]];
 G.~J.~Olmo,  AIP Conf.\ Proc.\  {\bf 1241}, 1100-1107 (2010), [arXiv:0910.3734 [gr-qc]].

\bibitem{Novello-2008}
M.Novello and S.E. Perez Bergliaffa, {\it Phys.Rep.} 463, 127-213 (2008);

\bibitem{Koi-2010}
T.Koivisto, Phys. Rev. {\bf D 82}, 044022 (2010).

\bibitem{BO-2010}
C. Barragan and G. J. Olmo, Phys. Rev. {\bf D 82}, 084015 (2010).


\bibitem{OSAT}
 G.~J.~Olmo, H.~Sanchis-Alepuz, S.~Tripathi,
  Phys.\ Rev.\  {\bf D80}, 024013 (2009).

\bibitem{Vitagliano:2010pq}
H. A. Buchdahl, J. Phys. A: Math. Gen. 12, 1235 (1979);
  V.~Vitagliano, T.~P.~Sotiriou and S.~Liberati,
  Phys.\ Rev.\  D {\bf 82}, 084007 (2010).

\bibitem{Vollick:2006uq}
  D.~N.~Vollick, 
  arXiv:gr-qc/0601016.

\bibitem{Olmo-2011}
G. J. Olmo, {\it Palatini actions and quantum gravity phenomenology}, [arXiv:1101.2841 [gr-qc]].

\bibitem{DSR-RG}
J.~Magueijo and L.~Smolin, Phys.Rev.Lett.{\bf 88}, 190403(2002);  Phys.Rev. D{\bf 67}, 044017(2003);
J.Magueijo and L.Smolin, Class. Quant. Grav.{\bf 21}, 1725 (2004).


\end{thebibliography}
\end{document}